\documentclass[preprints,article,accept,oneauthor,pdftex]{mdpi}
\firstpage{1} 
\pubvolume{1}
\issuenum{1}
\articlenumber{0}
\pubyear{2021}
\copyrightyear{2020}
\datereceived{} 
\dateaccepted{} 
\datepublished{} 
\hreflink{https://doi.org/} 

\Title{Retrospective of Research with Reinhard Schlickeiser}

\TitleCitation{Research with Reinhard Schlickeiser}

\Author{Charles Dermer$^{1}$}

\AuthorCitation{Dermer, C. D. }

\address{%
$^{1}$ Alexandria, Virginia, USA; charlesdermer@outlook.com; CharlesDermer.com\\}

\abstract{}
\abstract{I give a summary of my research with Prof.\ Dr.\ Reinhard Schlickeiser. It mainly took place in the 1990s, when the {\it Compton Gamma Ray Observatory} was giving a startling new view of the $\gamma$-ray sky. Our work focused on particle acceleration and radiation processes in the jets of active galactic nuclei (AGN). We pioneered the external Compton scattering model of blazars, where photons from outside the jet are intercepted and scattered to high energies by the radio-emitting electrons within the jet. Although we originally focused on external photons from the accretion disks of AGN, this process has been extended to include a range of external photon sources, and is now established as the primary source of $\gamma$ rays from flat spectrum radio quasars. }

\keyword{gamma-ray astronomy; plasma astrophysics; black-hole jets; cosmic rays; active galactic nuclei}

\begin{document}
\begin{quote}
{\it No one remembers the former generations,\\
    and even those yet to come\\
will not be remembered\\
    by those who follow them.\\
}
     \end{quote}
     \vskip-0.2in
      \hskip1.0in {\small Ecclesiastes 1:11, New International Version}

\section{Introduction}

I am extremely pleased to wish a happy 70th birthday to Reinhard Schlickeiser. Reinhard had a profound effect on my research in $\gamma$-ray and cosmic-ray astrophysics. He has been a strong advocate for many researchers in high-energy, cosmic-ray and plasma astrophysics, none more so than myself. His influence has improved our understanding of the cosmic-ray and $\gamma$-ray sky in innumerable ways. Here I will focus  on those few subjects on which we collaborated. 

It is already six years since I retired from the US Naval Research Laboratory, so I unfortunately do not have new research to present. I think, however, that it is appropriate on this occasion to reflect back to the heady days of the 1990s when space-based $\gamma$-ray astronomy was emerging from its beginnings, and ground-based $\gamma$-ray astronomy was being born. During this period of confusion,  Reinhard and I were in the middle of the mix, trying to understand the meaning of the new results coming in. I only hope that  younger researchers are able to participate in an excitement similar to what we encountered as the $\gamma$-ray sky was being unveiled.

 I began this article with a dark epigraph from the Book of Ecclesiastes. This came to mind when, years ago, I encountered a younger researcher and rising star in our field who seemed not to know who Reuven Ramaty was. Reuven supervised my first postdoc at NASA's Goddard Space Flight Center from 1984--1986. An energetic and driven scientist (Reinhard's colleague Ian Lerche referred to Ramaty as an "astrophysical engineer"), Reuven was a guiding light to myself and also, I believe, Reinhard. He had a Solar Satellite named after him, namely the {\it Reuven Ramaty High Energy Solar Spectroscopic Imager}. (Reuven died in 2001.) It dismayed me to find that Ramaty was almost forgotten, but this is in line with the presumption that everything before one's PhD is past history. Well, Ecclesiastes also says that {\it There is nothing new under the Sun}. But I'll be writing mainly about distant galaxies and here, at least, are many new things.  
 
 \section{Meeting Reinhard}
 
 Though I have met hundreds of scientists in my life, the memory of my initial meeting with Reinhard remains vivid. It was at a 1986 NORDITA workshop in Copenhagen organized by Roland Svensson on high-energy astrophysics, in the same room that Neils Bohr had held his famous meetings on quantum mechanics. One of the interesting questions of the time was the reason for the rather narrow range of the 2--10 keV X-ray spectral indices observed from AGN, which surely reflected a feature of particle acceleration. Reinhard's reputation as an expert in particle acceleration, particularly second-order processes, was already well known \cite{Schlickeiser(1984),Schlickeiser(1985)}. His approach was more analytical than numerical, so I felt a kinship and we struck up a correspondence.
 
Reinhard kindly invited to visit the MPI f\"ur Radioastronomie in Bonn where he was then working. We first treated a focused problem on radiation physics in order to see how well we could work together. I had already known Apostolos Mastichiadis since 1985, and we became interested in the work of his PhD thesis on the triplet pair production process where ultrarelativistic electrons pair-produce by interacting with photons ($e\gamma\rightarrow  ee^+e^-$) \cite{MMB86,MBM87,Mas91}. A full numerical treatment requires the use of the extremely complex cross sections derived by Haug \cite{Haug75,Haug81}.

Using  analytic arguments, we derived a simple expression for the energy-loss rate of ultra-relativistic electrons with Lorentz $\gamma$ interacting with isotropically distributed soft photons with dimensionless energy $\epsilon = E_{ph}/m_ec^2\ll 1$ and number density $n_{ph}$, given by
\begin{equation}
    {d\gamma\over dt} \cong  {4\over 3}\alpha_f c \sigma_{\rm T} n_{ph}  \;\large ({\gamma\over \epsilon}\large )^{{1/2}}\,
\end{equation}
\cite{DS91}. Compared to the exact numerical results of Mastichiadis and colleagues \cite{MMB86,MBM87}, this expression is accurate to a factor of 3 over more than 5 orders of magnitude in the variable $\epsilon \gamma$ characterizing the invariant energy of the interaction. (Here $\alpha_f \cong 1/137$ is the fine structure constant and $\sigma_{\rm T} =6.6525\times 10^{-25}$ cm$^{2}$ is the Thomson cross section.) The mean Lorentz factor of the produced electrons or positrons is $\langle  \gamma_1\rangle \cong (4/3)(\gamma/\epsilon)^{1/2}$. The paper goes on to examine the importance of this process in pair cascades initiated in extreme environments, such as Cygnus X-3 \cite{Sch89} and SN1987A \cite{SS91}, again comparing with detailed numerical results \cite{Mas91}.

\section{Enter the Compton Gamma-Ray Observatory}
 
 By 1990, space-based $\gamma$-ray astronomy had completed its pioneering phase, with three satellites---OSO-3, SAS-2, and COS-B---having detected $\gamma$ rays with energies $\gtrsim 30$ MeV  (see \cite{Der13} for a more detailed summary). OSO-3 and SAS-2 had discovered the $\gamma$-ray glow from the band of the Milky Way, a few $\gamma$-ray pulsars (Crab, Vela, and Geminga), and an isotropic $\gamma$-ray background. COS-B, the most advanced $\gamma$-ray telescope of the three, had reported 25 $\gamma$-ray sources, most of which were along the Galactic plane. But for the research that Reinhard and I were to be most interested in, the COS-B discovery of 3C 273, with emissions $> 100$ MeV, stood out. 
 
 Yet the appearance of the sole quasar 3C 273 on this list of sources did not give a clear indication of what was to come. 3C 273 is an unusual object in that, though radio-loud, it has an extraordinarily luminous accretion disk, suggesting that the emission might emanate directly from the disk itself. The 1983 review by Bassani \& Dean \cite{BD83} cited a handful of $\gamma$-ray AGN, but it was a mixed bag. These included the radio-quiet Seyfert galaxies NGC 4151 and MCG 6-11-11, the radio galaxy Centaurus A, the peculiar galaxy NGC 1275, and the quasar 3C 273. But only 3C 273 was detected at $\gtrsim 100$  MeV energies---by COS-B---whereas the others had detected emissions extending to only a few hundred keV. 

The Gamma-Ray Observatory, soon to be renamed the Compton Gamma-Ray Observatory (CGRO), was launched 5 April 1991. Its high-energy detector, the Energetic Gamma Ray Experiment Telescope (EGRET), was a spark chamber that tracked a $\gamma$ ray after it converted to an electron-positron pair by pair-production interactions with nuclei in the thin tantalum foils within the gas-filled spark chamber. It was sensitive to photons with energies between $\approx 30$ MeV -- 10 GeV. I was working as a research scientist at Rice University in Houston, Texas when it was launched. The CGRO researchers soon reported the detection of 3C 279, not, to our surprise, 3C 273. This indicated that the $\gamma$-ray emission from 3C 273 was variable, so must be associated with a rather compact emission region, because if it were from an extended region, light travel times from the different emission regions would erase the variability. 

I visited Reinhard at the MPIfR in August, 1991. By then we had learned that PKS 0528+134 had also been detected, so the race was on to determine what these sources had in common. I had little knowledge of radio-loud AGNs at that time, as my previous research had focused on gamma-ray bursts and relativistic thermal plasmas to account for 100 keV -- MeV emissions from sources such as Cyg X-1. I undertook a crash course to learn about radio-bright AGNs. The features they had in common was that they all contained relativistic jets that emitted intense radio radiation. They were also associated with apparent superluminal motions of the jet that occur when viewing the relativistically ejected blob slightly off-axis. Furthermore, they divided into the flat-spectrum radio quasars (FSRQs) with intense emission lines, indicating the presence of a bright nuclear accretion disk illuminating clouds of the broad-line region to make strong atomic-line radiations, and BL Lac objects (BLs), with weak or absent emission lines. Collectively, these sources are referred to as blazars.

Reviewing the literature, we focused on the work by Blandford and K\"onigl \cite{BK79}, which invoked relativistic jets of radio-emitting electrons to explain the radio-loud AGNs. Due to the Doppler boosting of the radiation, they are particularly luminous when viewed along the jet axis. Alan Marscher had also written an important paper \cite{Mar80} where he calculated the multi-wavelength spectra emitted by such jets. In his work, high-energy emissions were made by the synchrotron self-Compton process (SSC), where the radio-emitting electrons Compton scatter their own radio emission to high energies. Yet we could not make the SSC process account for the extraordinarily luminous $\gamma$-ray emissions of 3C 279 or PKS 0528+134 in either its energetic output, which dominated the spectral energy distribution of the detected radiation, or the photon frequency at the peak of its spectral energy distribution. This peak generally landed at hard X-ray range for standard parameters used to explain the radio emissions from these sources.

During the same period, there had been interest in understanding why the inferred Doppler factors $\Gamma$ of the relativistic jets were in the range from $\approx 5$--50. One idea was that relativistic electrons were slowed through Compton drag to reach terminal Lorentz factors in this range \cite{Phi87}.  The Compton drag was accomplished by scattering photons of the external radiation field. Melia and K\"onigl \cite{MK89}, in particular, made a detailed study where the accretion-disk photons were responsible for the relativistic drag on highly relativistic electrons traveling rectilinearly outward along the axis of the accretion disk. But there was a big problem in using this model to explain the radio emissions of blazars: electrons traveling in straight lines cannot make radio radiation, which requires tangled electrons in the outflowing blob.

Reinhard and I had the idea---which looks obvious in retrospect, as all good ideas should---that a proper model of the multi-wavelength emission, including the $\gamma$-ray emission, must involve a relativistic blob of outflowing plasma entraining relativistic electrons. In the stationary frame of this blob, the electrons have a tangled distribution to make  the radio and SSC radiation, but intercept external photons and scatter them through the Compton process to high energies to make the $\gamma$ rays. The effects of the Lorentz factor $\Gamma$ of the blob leaving the central nucleus combined with the individual Lorentz factors $\gamma$ of the electrons in the stationary frame of the blob produce intense $\gamma$-ray emissions.

Joined by Mastichiadis \cite{DSS92}, we worked out the physics of this system for photons entering directly from behind the jet, as would be the case for accretion disk photons when the blob far from the nucleus. We found that the $\gamma$-ray emission from disk photons entering the jet from behind would be most intense not when the blob was observed exactly along the jet axis, but rather at an angle $\theta_{pk} \cong \beta$, where $\beta = 1/\sqrt{1-1/\Gamma^2}$. This was precisely the angle at which such sources show largest apparent superluminal motion of their radio emissions! We wrote the paper and submitted it to the journal {\it Nature}, from which it was promptly rejected. I was extremely depressed and  ready to shelve the paper, but Reinhard encouraged me to resubmit it to another journal. It was accepted in {\it Astronomy \& Astrophysics Letters}, and has by now garnered some 500 citations.

Our paper \cite{DSS92} paved the way for future analyses, but of course did not settle all questions. Because of the kinematics when scattering accretion-disk photons entering from behind the jet, sources observed almost directly down the jet axis would be far less $\gamma$-ray luminous and show generally smaller superluminal speed. We speculated that BLs were those radio-loud AGNs observed along the jet axis. Though this generally agreed with the observed differences between FSRQs and BLs, the true picture has been shown to be more complicated. The $\gamma$ rays from BLs are primarily SSC emissions. Their weaker accretion disks and broad-line emissions make external Compton scattering, as the process is referred, much less important than the SSC process in BLs as compared to FSRQs.

The EGRET experiment on the Compton Gamma Ray Observatory was a breakthrough mission with many spectacular new discoveries, most notably the detection of beamed gamma-ray emission from extragalactic point sources at large redshifts. At such time, it is very importance to publish the first mostly correct theoretical explanation, and the passage of time has established external Compton process as the correct radiation mechanism. 

\section{Further $\gamma$-ray Blazar Studies}
 
Reinhard and his family visited me in Houston, TX, in the spring of 1992, and we worked out the detailed scattering kinematics and production spectra for this model, now taking into account the range of angles of photons from an extended accretion disk \cite{DS93}, using a standard Shakura-Sunyaev disk model \cite{SS73}. We also used the full Compton cross section, including both Compton scattering in the Thomson and KN regime, as Reinhard had shown that strong deviations from the Thomson regime already occur when the dimensionless photon energy $\epsilon^\prime = E^\prime/m_ec^2$ in the rest frame of the electron takes place even when $\epsilon^\prime \lesssim 0.1$ \cite{Sch79}. The major simplifying assumption was the head-on approximation, where the direction of the incident photon's direction is exactly opposite to the direction of motion of the electron. This assumption is perfectly justified for ultra-relativistic electrons.   

We fitted the $\gamma$-ray spectra of 3C 279, 3C 273 and Mkn 421, demonstrating that intense $\gamma$-ray emission was made by this process. We also calculated energy-loss rates of the electrons in external Compton scattering, which are important for dynamical treatments of these sources. Our paper \cite{DS93} became one of the "legacy" papers of the blazar Compton era, having now received over 700 citations.
More important than citations is that this model became the standard starting point with which to model high-energy emission from blazars. Subsequent improvements to the model invoked external photons from broad-line region clouds \cite{SBR94} and infrared emission from the cool dusty infrared-emitting torus \cite{Bla00,APS02}.

Excited by the blazar results from CGRO, we wrote a {\it Science} paper in 1992 to promote the new class of $\gamma$-ray blazars \cite{DS92}. The number of blazar detections with EGRET had grown to 14 by May of 1992, with a mixture of superluminal sources, FSRQs, and BLs. The BLs numbered 4 out of the 14 sources, and would be found to be the most important extragalactic source class detected by the ground-based air Cherenkov telescopes. The pioneering Whipple telescope had confidently detected TeV radiation from the Crab nebula only in 1989 \cite{Wee89}, establishing the superiority of the imaging approach after 2 decades of confusing results with the "on-off" technique. In 1992, the first high-confidence detection of an extragalactic source, Markarian 421, was made with Whipple \cite{Pun92}, and the field of $\gamma$-ray blazars studies, born in 1991 and 1992, was in full throttle.

I visited Reinhard several times at the MPIfR throughout the 1990s, and we continued to work out the radiation physics of blazars. One of the problems we examined \cite{DS94} was the location of the emission sites of blazars, finding that the pair-production process ($\gamma\gamma\rightarrow e^+e^{-}$) was particularly important for TeV $\gamma$ rays in FSRQs as compared to the BL Lac objects, making FSRQs potentially less luminous TeV $\gamma$-ray sources. With Steve Sturner, who was then my postdoc at the NRL, we completed a long study in 1997 \cite{DSS97} of the radiation physics in the AGN jets. 

Reinhard and I returned later for a final detailed look \cite{DS02} at radiation processes in blazar jets, published in 2002. We had been treating the physics of the scattering by transforming the external photon field into the rest frame of the blob, doing the scattering in the blob frame, and then transforming back to the observer frame. Georganopoulos, Kirk, and Mastichiadis \cite{GKM01} showed that it was much simpler for scattering calculations, particularly when using the full Compton cross section that includes  Klein-Nishina effects, to instead transform the electron spectrum directly to the observer frame and do the scattering in this frame. Except for dynamical calculations which required electron energy-loss rates  in the blob frame, this approach has proven more convenient.

In the {\it Science} paper \cite{DS92}, we also performed a first crude calculation of the blazar contribution to the (apparently) diffuse extragalactic $\gamma$-ray background. We showed that blazars are likely to make up a significant if not dominant part of this emission. This has been confirmed with the {\it Fermi} Gamma-ray Large Area Space Telescope \cite{Abd10,Der13}, although cosmic-ray induced emissions from star-forming
galaxies are also significant, as well as, potentially, a dark-matter contribution.

One of the happiest outcomes of my visits to Bonn was meeting Reinhard's extremely talented PhD student, Markus B\"ottcher. Markus and I published a study \cite{BD95} in 1995 of reverberation mapping effects on  $\gamma$-ray  photons that pair-produce in response to X-ray variability of the accretion disk. We later worked together to write many papers on radiation physics \cite{FDB08} (with Dr.\ Justin Finke, who was Markus's PhD student at Ohio University and later my postdoc at NRL) of and evolutionary scenarios \cite{BD02} for blazars, and the physics of micro-quasars \cite{DB06} and gamma-ray bursts \cite{BD98}. Dr. B\"ottcher has gone on to have a successful academic career, first as a professor at Ohio University in Athens, Ohio, and currently as Professor and Chair of Astrophysics and Space Physics, North-West University, Potchefstroom, South Africa.

\section{Diverging Pathways}

By the beginning of the first decade of the new century, our interests were starting to diverge. We completed a study of particle acceleration through magnetic turbulence in 2000 \cite{SD00}. Reinhard had always been interested in the fundamental physics of astrophysical plasmas and cosmic rays, which culminated in his magnum opus {\it Cosmic-Ray Astrophysics} \cite{Sch02}. Although Reinhard encouraged me to continue in this direction and look, for example, at the physics of dusty plasmas that had then become of great interest, my focus had shifted to the ultra-high energy sky that was being unveiled by the {\it Auger Observatory} and {\it IceCube}. I had also set my sights on the successor to CGRO,  the {\it Fermi}  Gamma-ray Large Area Space Telescope (GLAST), which was  launched in 2008.

Our last papers together \cite{aha09,Hess2011} were on joint observational campaigns with  {\it Fermi} and the High-Energy Stereoscopic Imager {\it HESS} in the late 2000s, where we provided theoretical interpretations for the results of these campaigns. But by then my research, especially with Armen Atoyan \cite{AD01,AD03} (now professor at Concordia University in Montreal, Canada), and with Soebur Razzaque (currently professor at the University of Johannesburg in South Africa) and Justin Finke, now staff scientist at NRL \cite{FRD10}, had taken center stage. With Professor Govind Menon of Troy University, Troy, Alabama, I summarized my black-hole research in a 2009 book-length monograph \cite{DM09}. Although I visited Reinhard at the Ruhr University in Bochum after he moved there in 1998 from the MPIfR, my motivation had flagged. The dogged pursuit of physical truth (the naive reason I went into physics) had been replaced by the dogged pursuit for external funding. Seeing my productivity decline, and meeting institutional difficulties on travel and visitors (this criticism of NRL is overshadowed by my gratitude for the generous support that was provided to me to perform basic research over more than two decades), I cashed in my chips and retired.

\section{Reinhard's Wisdom}

The most important thing I learned from Reinhard, not only by his words but by his actions, is the importance of collegiality in science. Many of us, and certainly I speak for myself, chose a path in science by virtue of being asocial, introverted, and hard working. But to only be a hard worker can threaten colleagues who are not so hard working, so the human side of science is just as important for all of us who find ourselves not to be the solitary geniuses that had inspired us (in my case, Newton and Cavendish). Rearranging the American proverb, {\it Hard work beats talent when talent doesn't work hard}, for the academic world it becomes  {\it Collegiality beats hard work when hard workers are not collegial}. 

This mirrors the apocryphal story, complete with cultural stereotypes, of the German and Frenchman who were to present their research for funding to a European Commission. The German stays up all night to perfect his presentation, but when he meets the Frenchman before the presentations, he is told that it is all settled: the Frenchman had wined and dined the head of the Commission and was assured that he, the Frenchman, would be selected. 

I also remember well Reinhard's allegory for conferences and workshops: they are like a fish market. A conference is the fishmonger showing his fish from a distance, all beautiful, shiny and fresh, whereas a workshop is when you get nearer to the market and smell that the fish are starting to rot and do not look so good on closer inspection. A conference is to show the great side of your work, and a workshop is to examine its shortcomings.

A final bit of wisdom I learned from Reinhard is the importance, at least for astrophysical theorists who try to understand observations, of being friends with the observers. They understand the pitfalls of their detectors and limitations of the data, and can give advanced notice of new results. I had my sources at NRL, NASA's GSFC and MSFC, and Reinhard had his at MPIfR and MPI f\"ur Extraterrestrische Physik  in Garching. They were anxious to share their new discoveries and see what the theorists might think, especially when they have results that are as yet unexplained.

My visits to Bonn and Bochum, the cigars and lively conversations, as well as my visits to Reinhard's house to meet his wife Wernhild and children Christina and Frank, will remain treasured memories. Happy 70$^{th}$ birthday, Reinhard, and may you have many, many more!


\linenumbers
\switchcolumn




\end{paracol}
\reftitle{References}



\begin{thebibliography}{999}

\bibitem[Abdo et al.(2010)]{Abd10} Abdo, A.~A. and 183 colleagues. Spectrum of the Isotropic Diffuse Gamma-Ray Emission Derived from First-Year Fermi Large Area Telescope Data. {\em PRL} {\bf 2010}, {\em 104}, 101101--101104.
\bibitem[Aharonian et al.(2009)]{aha09} Aharonian, F. and 325 colleagues. Simultaneous Observations of PKS 2155-304 with HESS, Fermi, RXTE, and Atom: Spectral Energy Distributions and Variability in a Low State. {\em ApJ} {\bf 2009}, {\em 696}, L150--L155.
\bibitem[Arbeiter et al.(2002)]{APS02} Arbeiter, C., Pohl, M., Schlickeiser, R. The influence of dust on the inverse Compton emission from jets in Active Galactic Nuclei.\ {\em A\&A} {\bf 2002}, {\em 386}, 415--426.

\bibitem[Atoyan and Dermer(2001)]{AD01} Atoyan, A.; Dermer, C.~D.  High-Energy Neutrinos from Photomeson Processes in Blazars. {\em PRL} {\bf 2001}, {\em 87}, 221102-1 -- 221102-4.
\bibitem[Atoyan and Dermer(2003)]{AD03} Atoyan, A.~M.; Dermer, C.~D. Neutral Beams from Blazar Jets. {\em ApJ} {\bf 2003}, {\em 586}, 79--96.


\bibitem[Bassani and Dean(1983)]{BD83} Bassani, L.; Dean, A.~J.\ 1983. Extragalactic {\ensuremath{\gamma}}-ray astronomy. {\em Space Science Reviews} {\bf 1983}, {\em 35}, 367--398.

\bibitem[Blandford and K{\"o}nigl(1979)]{BK79} Blandford, R.~D., K{\"o}nigl, A. Relativistic jets as compact radio sources. {\em ApJ} {\bf 1979}, {\em 232}, 34--48. 
\bibitem[B{\l}a{\.z}ejowski et al.(2000)]{Bla00} B{\l}a{\.z}ejowski; M., Sikora; M.; Moderski, R.; Madejski, G.~M. Comptonization of Infrared Radiation from Hot Dust by Relativistic Jets in Quasars. {\em ApJ}  {\bf 2000}, {\em 545}, 107--116. 
\bibitem[Boettcher and Dermer(1995)]{BD95} B\"ottcher, M.; Dermer, C.~D. Reverberation mapping of the central regions of active galactic nuclei using high-energy {\ensuremath{\gamma}}-ray observations. {\em A\&A} {\bf1995}, {\em 302}, 37--44.
\bibitem[B{\"o}ttcher and Dermer(2002)]{BD02} B{\"o}ttcher, M.; Dermer, C.~D. An Evolutionary Scenario for Blazar Unification. {\em ApJ} {\bf 2002}, {\em 564}, 86--91.
\bibitem[B{\"o}ttcher and Dermer(1998)]{BD98} B{\"o}ttcher, M.; Dermer, C.~D. High-energy Gamma Rays from Ultra-high-energy Cosmic-Ray Protons in Gamma-Ray Bursts. {\em ApJ} {\bf 1998}, {\em 499}, L131---L134.



\bibitem[Dermer(2013)]{Der13} Dermer, C.~D.\ 2013.\ Sources of GeV Photons and the Fermi Results.\ In {\em Astrophysics at Very High Energies}, Walter, R., T\"urler, M., Eds.; Saas-Fee Advanced Course 40, 2013, 225--356. 
\bibitem[Dermer and B{\"o}ttcher(2006)]{DB06} Dermer, C.~D.; B{\"o}ttcher, M. Gamma Rays from Compton Scattering in the Jets of Microquasars: Application to LS 5039. {\em ApJ} {\bf 2006}, {\em 643}, 1081--1097.
\bibitem[Dermer and Menon(2009)]{DM09} Dermer, C.~D.; Menon, G. \textit{High Energy Radiation from Black Holes: Gamma Rays, Cosmic Rays, and Neutrinos.} Princeton Univerisity Press: Princeton, NJ, 2009.

\bibitem[Dermer and Schlickeiser(1991)]{DS91} Dermer, C.~D.; Schlickeiser, R. Effects of triplet pair production on ultrarelativistic electrons in a soft photon field. {\em A\&A} {\bf1991}, {\em 252}, 414--420.
\bibitem[Dermer and Schlickeiser(1992)]{DS92} Dermer, C.~D., Schlickeiser, R.\ Quasars, Blazars, and Gamma Rays.  {\em Science} {\bf 1992}, {\em 257}, 1642--1647.
\bibitem[Dermer and Schlickeiser(1993)]{DS93} Dermer, C.~D.; Schlickeiser, R. Model for the High-Energy Emission from Blazars. {\em ApJ} {\bf 1993}, {\em 416}, 458--484.
\bibitem[Dermer and Schlickeiser(1994)]{DS94} Dermer, C.~D.; Schlickeiser, R. On the Location of the Acceleration and Emission Sites in Gamma-Ray Blazars. {\em ApJS} {\bf 1994}, {\em 90}, 945--948.
\bibitem[Dermer and Schlickeiser(2002)]{DS02} Dermer, C.~D.; Schlickeiser, R. Transformation Properties of External Radiation Fields, Energy-Loss Rates and Scattered Spectra, and a Model for Blazar Variability. {\bf 2002}, {\em 575}, 667--686.
\bibitem[Dermer et al.(1992)]{DSS92} Dermer, C.~D.; Schlickeiser, R.; Mastichiadis, A. High-energy gamma radiation from extragalactic radio sources. {\em A\&A} {\bf 1992}, {\em 256}, L27--L30.
\bibitem[Dermer et al.(1997)]{DSS97} Dermer, C.~D.; Sturner, S.~J.; Schlickeiser, R. Nonthermal Compton and Synchrotron Processes in the Jets of Active Galactic Nuclei. {\em ApJS} {\bf 1997}, {\em 109}, 103--137.

\bibitem[Finke et al.(2008)]{FDB08} Finke, J.~D.; Dermer, C.~D.; B{\"o}ttcher, M.  Synchrotron Self-Compton Analysis of TeV X-Ray-Selected BL Lacertae Objects. {\em ApJ}  {\bf 2008}, {\em 686}, 181--194.
\bibitem[Finke et al.(2010)]{FRD10} Finke, J.~D.; Razzaque, S.; Dermer, C.~D. Modeling the Extragalactic Background Light from Stars and Dust. {\em ApJ} {\bf 2010}, {\em 712}, 238--249.
\bibitem[Georganopoulos et al.(2001)]{GKM01} Georganopoulos, M.; Kirk, J.~G.; Mastichiadis, A. The Beaming Pattern and Spectrum of Radiation from Inverse Compton Scattering in Blazars. {\em ApJ} {2001}, {\em 561}, 111--117. 

\bibitem[Haug(1975)]{Haug75} Haug, E. Bremsstrahlung and pair production in the field of free electrons. {\em Zeitschrift Naturforschung} {\bf 1975}, {\em A 30}, 1099--1113.
\bibitem[Haug(1981)]{Haug81} Haug, E.  Simple Analytic Expressions for the Total Cross Section for {\ensuremath{\gamma}}-e Pair Production.\ {\em Zeitschrift Naturforschung} {\bf 1981}, {\em A 36}, 413--414.
\bibitem[H.~E.~S.~S. Collaboration et al.(2011)]{Hess2011} {\it HESS} and {\it Fermi} Collaboration and 348 colleagues. Simultaneous multi-wavelength campaign on PKS 2005-489 in a high state. {\it A\&A} {bf 2011},  {\em 533}, 110-118.

\bibitem[Marscher(1980)]{Mar80} Marscher, A.~P. Relativistic jets and the continuum emission in QSOs. {\em ApJ} {\bf 1980}, {\em 235}, 386--391.

\bibitem[Mastichiadis et al.(1986)]{MMB86} Mastichiadis, A.; Marscher, A.P.; Brecher, K. Electron-Positron Pair Production by Ultrarelativistic Electrons in a Soft Photon Field {\em ApJ} {\bf 1986}, {\em 300}, 178--189.
\bibitem[Mastichiadis et al.(1987)]{MBM87} Mastichiadis, A.; Brecher, K.; Marscher, A.~P. Electromagnetic Cascades in the Magnetosphere of a Very Young Pulsar: A Model for the Positron Production near the Galactic Center. {\em ApJ} {\bf 1987}, {\em 314}, 88--94.

\bibitem[Mastichiadis(1991)]{Mas91} Mastichiadis, A. Relativistic electrons in photon fields --- Effects of triplet pair production on inverse Compton gamma-ray spectra. {\em MNRAS} {\bf 1991}, {\em 253}, 235--244.


\bibitem[Melia and K\"onigl(1989)]{MK89} Melia, F.; K\"onigl, A. The Radiative Deceleration of Relativistic Jets in Active Galactic Nuclei. {\em ApJ} {\bf 1989}, {\em 340}, 162--183.

\bibitem[Phinney(1987)]{Phi87} Phinney, E.~S. How fast can a blob go? In {\em Superluminal Radio Sources}, Zensus, J. A., Pearson, T.J., Eds.; Cambridge University Press, Cambridge, UK, 1987, pp. 301--305.
\bibitem[Punch et al.(1992)]{Pun92} Punch, M. and 23 colleagues. Detection of TeV photons from the active galaxy Markarian 421. {\em Nature} {\bf 1992} {\em 358}, 477--478.

\bibitem[Schlickeiser(1979)]{Sch79} Schlickeiser, R. Astrophysical gamma-ray production by inverse Compton interactions of relativistic electrons. {\em ApJ} {\bf 1979} {\em 233}, 294--301
\bibitem[Schlickeiser(1984)]{Schlickeiser(1984)} Schlickeiser, R. An explanation of abrupt cutoffs in the optical-infrared spectra of non-thermal sources. A new pile-up mechanism for relativistic electronspectra. {\em A\&A} {\bf 1984}, {\em 136}, 227--236.
\bibitem[Schlickeiser(1985)]{Schlickeiser(1985)} Schlickeiser, R. A viable mechanism to establish relativistic thermal particle distribution functions in cosmic sources. {\em A\&A} {\bf 1985}, {\em 143}, 431-434.
\bibitem[Schlickeiser(1989)]{Sch89} Schlickeiser, R.\ 1989.\ PeV inverse Compton gamma rays from Cygnus X-3. {\em A\&A} {\bf 1989}, {\em 213}, L23--L25.
\bibitem[Schlickeiser(2002)]{Sch02} Schlickeiser, R. \textit{Cosmic Ray Astrophysics}, Springer: Berlin, 2002.
\bibitem[Schlickeiser and Dermer(2000)]{SD00} Schlickeiser, R.; Dermer, C.~D. Proton and electron acceleration through magnetic turbulence in relativistic outflows. {\em A\&A} {\bf 2000}, 360, 789--794.
\bibitem[Schlickeiser and Stanev(1991)]{SS91} Schlickeiser, R.; Stanev, T.\ 1991.\ Origin of the burst of TeV gamma-rays from SN 1987A. {\em A\&A} {\bf 1991}, {\em 243}, L1--L4.

\bibitem[Shakura and Sunyaev(1973)]{SS73} Shakura, N.~I.; Sunyaev, R.~A.\ 1973. Black holes in binary systems. Observational appearance. {\em A\&A} {\bf 1973}, {\em 500}, 33--51.

\bibitem[Sikora et al.(1994)]{SBR94} Sikora, M.; Begelman, M.~C.; Rees, M.~J. Comptonization of Diffuse Ambient Radiation by a Relativistic Jet: The Source of Gamma Rays from Blazars? {\em ApJ} {\bf 1994}, {\em 421}, 153--?. 

\bibitem[Weekes et al.(1989)]{Wee89} Weekes, T.~C. and 11 colleagues. Observation of TeV Gamma Rays from the Crab Nebula Using the Atmospheric Cerenkov Imaging Technique. {\em ApJ} {\bf 1989}, {\em 342}, 379--395.

\end{thebibliography}


\end{document}